# Noise Speech wavelet analyzing in special time ranges

Amin Daneshmand Malayeri
Computer Engineering Department
Hamedan University of Technology, Young Researchers Club of Malayer Azad University
Hamedan , Iran
E-mail : amin.daneshmand@gmail.com

*Abstract*— speech analyzing in special periods of time has been presented in this paper. One of the most important periods in signal processing is near to Zero. By this paper, we analyze noise speech signals when these signals are near to Zero. Our strategy is defining some subfunctions and compress histograms when a noise speech signal is in a special period. It can be so useful for wavelet signal processing and spoken systems analyzing.

## I. INTRODUCTION

It is very important that we can review some treatments of functions in some special periods of time. One of these important matters occur in signal processing, especially in some wavelet-based signals. Removing noise components is based on the observation that in many signals, energy is mostly concentrated in a small number of wavelet dimensions [1].

In noise speech studies, when some factors in signal equation is near to Zero, we have interesting results for signal processing by some factors. Our vision is defining some coefficients for noise speech signal equations and limit them to a period near to Zero.

The study in [2] showed that when noise dominates the observed data, the universal threshold method performs well and when the underlying signal dominates the observed data, the SURE (Stein's Unbiased Risk Estimator) method [3,4] performs better than the universal method. This observation led to the heuristic SURE method, which selects either the universal threshold or the SURE threshold according to a test that finds out which of the noise or the underlying signal dominates the observed data [4]. Mini-Max threshold [5] is another method for estimating the appropriate threshold value and is obtained very similar to the one obtained in Donoho's method.

After the threshold estimation, a thresholding function should be determined. Using this function noisy wavelet coefficients are compared with the threshold value in order to determine which part of them should be modified. common thresholding functions are hard thresholding functions , soft thresholding function and semi-soft thresholding function [6-10].

We introduce a mathematical-based function in noise speech by analyzing some factors when the function is near to Zero.

At this state, the speech function and noise speech function diagrams will be linear and very similar.

In telecommunication systems, when a noise speech occurs, you can control it in small scale of time ranges. These ranges is near to Zero, in very short periods.

By mathematical operators, we can estimate these ranges. Our studies show that when noise speech and estimated noise signals are very near to Zero, both of them can cover themselves. Propagation of points in noise speech and estimated noise signals has been decreased and data transferring between these points can be make high accuracy.

## II. BASIC SPEECH FUNCTIONS

A wavelet based speech enhancement system, is composed of five steps:
- windowing and overlapping
- wavelet packet decomposition
- thresholding or filtering
- wavelet packet reconstruction
- adding and overlapping

some of the thresholding methods are listed in the following:
- Hard threshold
- Soft threshold
- Semi Hard threshold
- Semi Soft threshold
- Quintile based threshold

Hard threshold will be discussed in the following [11]. The standard form of Hard thresholding has an input-output characteristics that is drawn by solid line in Fig.1 .

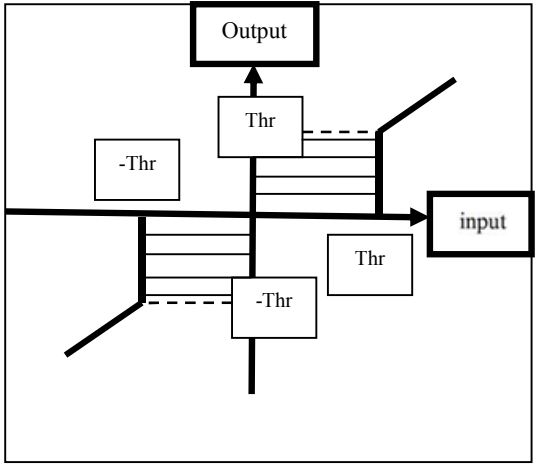

Figure 1. schematic for Input-Output Hard thresholding

The Hard threshold is defined using (1):

$$Thr_H(w_s^D, T) = \begin{cases} 0 & |w_s^D| < T \\ w_s^D & |w_s^D| \geq T \end{cases} \quad (1)$$

In order to use Hard threshold for noise reduction, we have to set a suitable tradeoff between the remained background noise and the distortion in the enhanced speech signals.

### III. SPEECH ANALYZING IN THE RANGE OF $[T_1, T_2]$

Speech processing area contains a wide range of applications on the speech signals such as speech enhancement, speech recognition, speaker recognition, speaker validation, spoken dialog systems, text to speech systems and voiced/unvoiced decoder. Although information theory concepts may be used in almost all mentioned fields but they are mainly used in the last three applications [12,13].

In the area of blind source separation , the mutual information [18] has been widely used. The blind source separation problem has been discussed in many references [14-18]. Suppose there are a number of speakers talking at the same time. Using microphones placed at different locations that are able to record mixtures of the signals coming from all the speakers. The objective in the blind source separation problem is to recover the original speech patterns from the linear mixtures. Here , "blind" implies that neither the mixing coefficients nor the probability distributions of the original sources are known.

In the context of voiced / unvoiced decoding, the entropy measure [14] has been used for achieving a good performance in the wavelet based speech enhancement systems .

In order to remove the noisy coefficients with low distortion in the enhanced speech signal, the value of threshold has to be different in the voiced and unvoiced frames. The value of the threshold in the unvoiced frames is smaller than it in the voiced frames.

The entropy measure is used to detect voiced / unvoiced frames. The entropy of the coefficients in the unvoiced frames is less than a predefined threshold. So, by selecting a suitable value for the threshold, a good performance could be achieved for its decoders.

The key feature that the problem solver has to explore is the statistical independence in the components of the original vector. It means that the problem solver has to impose as much statistical independence as possible on the individual components of the output vector.

The histogram of coefficients will be calculated for each sub-band. One of the critical parameter is the number of bins in the histogram. This parameter has been chosen equal to the square root of the number of samples divided by two. the probability distribution of the coefficients could be estimated .

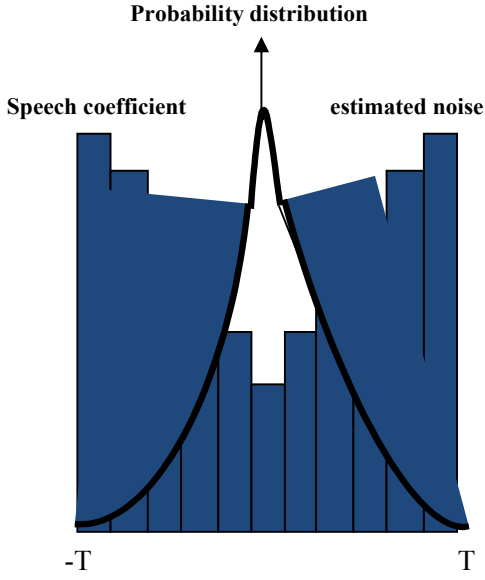

Figure 2. wavelet coefficient histogram

Our main equation for reviewing behaviour of the noise speech is defined as :

$$D(p \parallel q) = \sum_{i=1}^{M} p(x_i) \log\left(\frac{p(x_i)}{q(x_i)}\right) \quad (2)$$

That D is the divergence between two functions p(x) and q(x) .

$$p(x_i) = \frac{\text{noise speech in range of } [T1, T2]}{\text{estimated noise in range of } [T1, T2]} \quad (3)$$

$$\sum_{i=T1}^{T2} P_{ns}(i) \log \frac{P_{ns}(i)}{P_n(i)} + \sum_{i=T1}^{T2} P_n(i) \log \frac{P_n(i)}{P_{ns}(i)} \approx 0 \quad (4)$$

$$\log \frac{P_{ns}(i)}{P_n(i)} = z \rightarrow \sum_{i=T1}^{T2} P_{ns}(i)(z) + \sum_{i=T1}^{T2} P_n(i)(-z) \approx 0 \quad (5)$$

$$\sum_{i=T1}^{T2} z [P_{ns}(i) - P_n(i)] \approx 0 \rightarrow P_{ns}(i) \approx P_n(i) \quad (6)$$

$$\lim_{i \to 0} \frac{P_{ns}(i)}{P_n(i)} = 1 \quad (7)$$

All of equations show that when a noise speech signal is near to estimated noise, they can be similar, especially near to Zero ranges.

The equation (7) showed that $P_{ns}(i)$ is near to $P_n(i)$, when both of them are near to Zero.

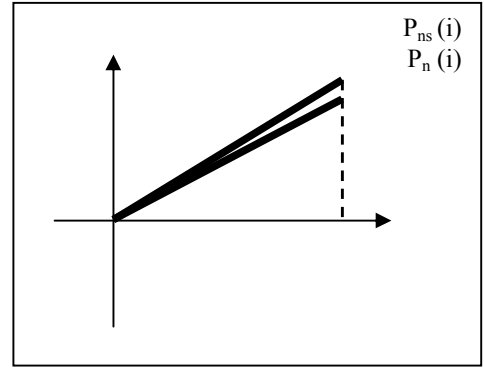

Figure3. $P_n(i)$ , $P_{ns}(i)$ graph near to 0

As $T_1, T_2$ are the number of bins in the histogram of noisy speech coefficients , we have to modify the hard thresholding method.
The coefficient with values in the range determined by $T_1, T_2$, marked as noise coefficients and will be removed. The other coefficients marked as clean speech coefficients and will be remained with no change.

### IV. CONCLUSION

in this paper, we introduce some speech and estimated noises in special ranges near to Zero. Our main purpose for optimization signals is making purity with the signals. We use from Mathematical analyzing for reviewing treatment of noise speech and estimated noise near to Zero ranges. In telecommunicating services, one of the most important matters for data transferring is decreasing noises and the other impurities for making the best path for goal signals.
By approximating histograms of noise speech and estimated noise, we can make them compatible and with the same treatment near to some ranges, especially in Zero ranges periods.
Also, we can improve wavelet-based signals near to Zero by analyzing them as describe in this paper.